\documentclass[12pt]{iopart}
\usepackage{graphicx}
\usepackage{dcolumn}
\usepackage{bm}
\usepackage{iopams}
\usepackage{color}

\newcommand{\beqar}{\begin{eqnarray}}
\newcommand{\eeqar}{\end{eqnarray}}
\newcommand{\bea}{\begin{eqnarray}}
\newcommand{\eea}{\end{eqnarray}}
\newcommand{\bcen}{\begin{center}}
\newcommand{\ecen}{\end{center}}

\newcommand{\bra}[1]{\left< #1 \right|}
\newcommand{\ket}[1]{\left| #1 \right>}

\newcommand{\f}[2]{\frac{#1}{#2}}
\renewcommand{\b}[1]{\left({#1}\right)}
\renewcommand{\v}[1]{\vec{#1}}
\newcommand{\pd}[2]{\frac {\partial #1}{\partial #2}}

\renewcommand{\sb}[1]{\left[{#1}\right]}
\newcommand{\mean}[1]{\langle {#1} \rangle}

\newcommand{\ra}{\rightarrow}

\begin{document}

\title[Quantum Signatures in the Quantum Carnot Cycle]{Quantum Signatures in the Quantum Carnot Cycle}

\author{Roie Dann}
\address{The Institute of Chemistry, The Hebrew University of Jerusalem, Jerusalem 9190401, Israel}
\address{Kavli Institute for Theoretical Physics, University of California, Santa Barbara, CA 93106, USA}
\ead{roie.dann@mail.huji.ac.il}
\author{Ronnie Kosloff}
\address{The Institute of Chemistry, The Hebrew University of Jerusalem, Jerusalem 9190401, Israel}
\address{Kavli Institute for Theoretical Physics, University of California, Santa Barbara, CA 93106, USA}
\ead{ronnie@fh.huji.ac.il}
\vspace{10pt}
\begin{indented}
\item[]August 2019
\end{indented}

\begin{abstract}
The Carnot cycle combines reversible isothermal and adiabatic strokes to obtain optimal efficiency, at the expense of a vanishing power output. Quantum Carnot-analog cycles are constructed and solved, operating irreversibly at non-vanishing power. Swift thermalization is obtained in the isotherms utilizing a shortcut to equilibrium protocols and the adiabats employ frictionless unitary shortcuts. The working medium in this study is composed of a particle in a driven harmonic trap. For this system, we solve the dynamics employing a generalized canonical state. Such a description incorporates both changes in energy and coherence. 
This allows comparing three types of Carnot-analog cycles, Carnot-shortcut, Endo-shortcut and Endo-global.
The Carnot-shortcut engine demonstrates the trade-off between power and efficiency. It posses a maximum in power, a minimum cycle time where it becomes a dissipator and for a diverging cycle time approaches the ideal Carnot efficiency.
The irreversibility of the cycle arises from non-adiabatic driving, which generates coherence. To study the role of coherence we compare the performance of the shortcut cycles, where coherence is limited to the interior of the strokes, with the Endo-global cycle where the coherence never vanishes. The Endo-global engine exhibits a quantum signature at a short cycle-time, manifested by a positive power output while the shortcut cycles become dissipators.
If energy is monitored the back action of the measurement causes dephasing and the power terminates. 

\end{abstract}

%
%
%
%
%
\break

\section{\label{sec:Intro}Introduction}

In 1824 Sadi Carnot envisioned a reversible reciprocating heat engine, composed of two adiabatic and two isothermal strokes. He argued that such a reversible engine is universal, and produces optimal work, depending only on the hot and cold bath temperatures, $T_h$ and $T_c$ \cite{carnot1824reflexions}. This cycle is termed Carnot cycle and serves as a template for all reversible engines. Generally, a heat engine transforms heat extracted from a hot bath, $Q_h$, to work $W$. According to the second law of thermodynamics this work is necessarily accompanied by a heat flow to the cold bath $Q_c =|W|-Q_h$. The efficiency of the energy transformation is defined by  $\eta=-W/Q_h$, which is bounded by the Carnot cycle efficiency $\eta<\eta_C=1-T_c/T_h$, for any heat cycle. This universal bound depends only on the bath temperatures, irrespective of the specific working medium. It translates a fundamental limitation, the second law of thermodynamics, to a practical operational limit applying to all engines. Optimal efficiency is achieved only for a reversible operation.

Practically, any finite power engine operates under irreversible conditions, which leads to a trade-off between efficiency and power. This relation has been extensively studied in the framework of finite-time thermodynamics \cite{salamon2001principles,andresen1984thermodynamics}. A prominent emerging result states that under endoreversible conditions the efficiency at maximum power is given by the Curzon-Ahlborn-Novikov efficiency $\eta_{CA} = 1-\sqrt{T_c/T_h}$ \cite{curzon1975efficiency,novikov1958efficiency}, which has been generalized for weak dissipation \cite{esposito2010quantum,esposito2010efficiency}.

In the quantum regime energy transfer is constrained as well by the second law of thermodynamics \cite{alicki1979quantum,kosloff2013quantum,vinjanampathy2016quantum,gemmer20044}. Implying that 
quantum heat engines are also bounded by the Carnot efficiency. Reversibility and optimal efficiency is obtained in the quantum adiabatic limit, requiring an infinite cycle-time $\tau_{cycle}$.

 In this paper we explore two quantum approaches to obtain non-vanishing power for a `Carnot-analog' engine. The first type is termed {\it{Carnot-shortcut}}, defined by the cycle parameters of the reversible Carnot cycle (bath temperatures and external parameters). At the switching corners between strokes,  the working medium is in equilibrium with the baths and out of equilibrium at intermediate times (Fig. \ref{fig:H_vs_omega} Panel (a)). Finite power is achieved by shortcut protocols to each stroke: Shortcuts To Adiabaticity (STA) \cite{chen2010fast} on the unitaries and Shortcut To Equilibrium (STE) \cite{dann2018shortcut} for the thermalization strokes.
 The cycle is performed by external driving of the system and  coupling/decoupling the Working Medium (WM) from the hot and cold baths. Typically, the WM Hamiltonian does not commute with itself at different times $\sb{\hat{H}_S\b t,\hat{H}_S \b{t'}}\neq 0$, leading to generation of coherence and an accompanied cost in work \cite{feldmann2003quantum,francica2019role,plastina2014irreversible}.
 
An alternative approach to obtain finite power is a {\it{Quantum Endoreversible}} cycle (second type of 'Carnot analog' engine), for which the WM is in a non-equilibrium state throughout the cycle. In the class of endoreversible cycles, we construct the {\em{Endo-shortcut}} and the {\em{Endo-global}} cycles, which are characterized by {\it{Branch}} and {\it{Global}} coherence operations, respectively. Branch coherence is restricted to the interior of the stroke, while at the cycle's four corners the coherence vanishes and the WM state is of a Gibbs form, with an internal temperature $T\neq T_{bath}$. In the global coherence operation the WM state exhibits coherence throughout the whole cycle. As a result, coherence generated in one stroke continues to play a role in the subsequent strokes. 

Our current aim is to explore the performance and characteristics of the Quantum Carnot-analog engines, i.e., the trade-off between power and efficiency and the role of quantum coherence in the engine operation. For this purpose, we employed an harmonic oscillator WM, allowing an explicit solution of the cycle dynamics. Recently, such a working medium has been employed to  experimentally realize quantum heat engines \cite{rossnagel2016single,von2019spin}.

Previous studies of the Carnot cycle lacked coherence. These studies consider a Hamiltonian that commutes with itself at all times \cite{geva1992quantum,wu2006performance}, or cycles performed in the adiabatic \cite{bender2000quantum,altintas2019comparison} or stochastic \cite{esposito2010quantum} limits. The relatively few studies on the quantum Carnot-analog cycle is in contrast to the popularity of the quantum Otto cycle.  Analysis of the Otto cycle has been a major source of insight on quantum reciprocating engines \cite{feldmann2003quantum,kosloff2010optimal,kosloff2017quantum,henrich2007quantum,rossnagel2016single,rossnagel2014nanoscale,abah2012single,del2014more,beau2016scaling,campisi2016power,ccakmak2019spin,abah2019shortcut,ccakmak2017irreversible,peterson2018experimental,camati2019coherence}. The vast difference in popularity between the quantum Otto and Carnot cycles arises from the difficulty to describe the open-system dynamics of non-adiabatically driven systems. Recent development of the Non-Adiabatic Master Equation (NAME) \cite{dann2018time} and the inertial theorem \cite{dann2018inertial} enable this study.

The analysis of the quantum Carnot-analog engine, demonstrates two roles of coherence in the energy representation. First, the control of coherence allows optimizing the efficiency for a finite cycle-time, minimizing quantum friction \cite{feldmann2003quantum}. Moreover, for very short cycle-times global coherence becomes crucial to obtain non-vanishing power. This phenomena is a quantum signature \cite{uzdin2015equivalence,klatzow2019experimental,miller2019work}, which can be unraveled from thermodynamics observables. We consider a thermodynamic quantum signature as a measurable thermodynamic quantity, such as work or efficiency, which indicates a unique quantum behaviour. Such behaviour, by definition, does not have a classical counterpart. Specifically, we find that under fast driving the coherent work extraction mechanism outperforms the stochastic work extraction ֿ\cite{uzdin2015equivalence}.   

The paper is organized as follows: We begin by addressing the general construction and features of the various Carnot analog cycles, Sec \ref{sec: Dynamics of Carnot-analog}. Section \ref{sec: Carnot shortcut cycle} then focuses on the Carnot-shortcut cycle, describing its construction and thermodynamics behaviour. In this framework, we discuss the role of coherence in the efficiency-power tradeoff, which naturally emerges in the performance of the Carnot-shortcut cycle. In Sec. \ref{sec: Endoreversible} we introduce two additional Carnot analog cycles, the endoreversible shortcut and global Carnot cycles. Following the description, we compare between branch and global coherence operation modes, associated with these two cycles. Such comparison, highlights the role of coherence in the fast driving regime, and illuminates the emergence of quantum thermodynamics signatures.  We conclude in Sec. \ref{sec:discussion} with a summary and compare the quantum Carnot and Otto engines.

\section{Dynamics of the Carnot-analog cycle}
\label{sec: Dynamics of Carnot-analog}
We choose a particle of mass $m$ confined by a varying Harmonic potential as the engine's working medium. The associated Hamiltonian reads 
\begin{equation}
    \hat{H}\b t=\f{\hat{P}^2}{2m}+\f{1}{2}m\omega^2\b t \hat{Q}^2~~,
    \label{eq:Ham}
\end{equation}
where $\hat{Q}$ and $\hat{P}$ are the position and momentum operators, and $\omega\b t$ is the oscillator frequency and the control parameter. The explicit time-dependence of $\omega\b t$ defines the cycle's protocol.

The dynamics during the cycle strokes is associated with a time-dependent completely positive map \cite{lindblad1976generators,gorini1976completely}, generated by
\begin{equation}
\f{d\hat{\rho}\b t}{dt} = {\cal{L}}\b t\hat{\rho}\b t~~,
\end{equation}
For the adiabatic strokes ${\cal{L}}\b t\hat{\rho}\b t=-i\sb{\hat{H}\b t,\hat{\rho}\b t}$, generating the unitary maps of the adiabatic expansion and compression ${\cal{U}}_{ch}$ and ${\cal{U}}_{hc}$.

In the isothermal-type strokes ${\cal{L}}\b t\hat{\rho}\b t=-i\sb{\hat{H}\b t,\hat{\rho}\b t} +{\cal{L}}_D\b t \hat{\rho}\b t$, where ${\cal{L}}_D\b t$ is an explicitly time-dependent Lindblidian \cite{dann2018time}, generating the
isotherms dynamical maps, ${\cal{U}}_{h}$ and ${\cal{U}}_{c}$. Theses strokes incorporate non-adiabatic driving protocols. 

The generators of the isotherms can be derived from first principles under the following restrictions: (i) Weak system-bath coupling;  (ii) Fast bath dynamics relative to the system  and driving \cite{dann2018time}; (iii) The protocol complies with the Inertial Theorem \cite{dann2018inertial}. Specifically, for the parametric harmonic oscillator, Eq. (\ref{eq:Ham}), the inertial theorem is satisfied in the limit of a slow change in the adiabatic parameter, $d\mu/dt\ra 0$, where $\mu=\dot{\omega}/\omega^2$ . The adiabatic parameter can be large, indicating fast driving relative to the Bohr frequencies, while varying slowly. The resulting master equation is consistent with the thermodynamic laws \cite{dann2018time,k329}.

The Carnot analog cycle is constructed by concatenating four strokes:
\begin{itemize}
    \item An expansion stroke (Fig. \ref{fig:H_vs_omega} Panel (a) corners $1\ra2$), decreasing the oscillator frequency.
while the system is coupled to a hot bath of a temperature $T_h$. This stroke is termed {\emph{open-expansion}}, referring to the fact that during this segment the working medium is an open quantum system, exchanging energy and entropy with the bath. The evolution of the WM is governed by the propagator ${\cal{U}}_h$.
\item {\emph{Adiabatic expansion}} stroke ($2\ra3$)  in which the system is isolated from the baths and the oscillator frequency is decreased. The stroke is associated with the propagator ${\cal{U}}_{ch}$. 
\item \emph{Open-compression} ($3\ra4$), the particle is brought in contact with the cold bath of temperature $T_c$ and `compressed' towards a higher frequency, associated with the propagator ${\cal{U}}_c$. 
\item {\emph{Adiabatic compression}} ($4\ra1$), the final stroke restores the system to its initial state 
while keeping the particle isolated from the baths. The stroke propagator is ${\cal{U}}_{hc}$.
\end{itemize}
 Formally, the complete cycle propagator can be decomposed to stroke propagators:
 \begin{equation}
   {\cal U}_{cyc}={\cal U}_{hc}{\cal U}_{c}{\cal U}_{ch}{\cal U}_{h}~~,
   \label{eq:cycle_prop}
 \end{equation}
Each propagator ${\cal{U}}_i$ is a completely positive map, associated with the cycle stroke. For a closed Lie algebra, the system dynamics can  be described within a restricted operator vector space. Within this framework, we can represent the state as a generalized Gibbs form, which is the maximum entropy state under the constraint imposed by the observables of the Lie algebra \cite{alhassid1978connection,jaynes1957information,Jaynes1957}. The generalized Gibbs structure is preserved under the full dynamics of the cycle, this property is termed canonical invariance \cite{rezek2006irreversible,dann2018shortcut,dann2018time}.
For the  analysis, we choose  operators motivated by thermodynamical insight: The instantaneous Hamiltonian and two additional operators to represent coherence the Lagrangian $\hat L \b t$ and ${\hat{C}}\b t$ (Cf. \ref{apsec:Endo-global cycle}). These operators do not commute with $\hat{H}\b t$,  together they define the coherence measure \cite{kosloff2017quantum}:
\begin{equation}
    {\it{Coh}}=\f{\sqrt{\mean{\hat{L}\b t}^2+\mean{\hat{C}\b t}^2}}{\hbar \omega \b t}~~.
    \label{eq:coherence}
\end{equation}
Repeated operation of the cycle propagator, Eq. (\ref{eq:cycle_prop}), converges to a unique limit cycle \cite{feldmann2004characteristics}. 
The thermodynamic analysis is then carried out on the limit cycle in terms of work $W$ and heat $Q$. External work preformed on the WM is
defined in terms of the instantaneous power, $\rm{tr}\b{\dot{\hat{H}}\hat{\rho}}$, \cite{alicki1979quantum}
\begin{equation}
    W=\int_0^t \langle\pd{\hat{H}\b{t'}}{t'}\rangle dt'~~,
\end{equation}
and heat is obtained from the first law of thermodynamics $E=W+Q$.

\section{Carnot-shortcut cycle}
\label{sec: Carnot shortcut cycle}
The Carnot-shortcut cycle is characterized by the four corners of the ideal Carnot cycle.  Utilizing shortcut protocols, the WM transitions between the thermal states associated with the cycle corners in finite time. Starting from a thermal state the non-adiabatic protocols induce coherence generation and excitations leading to a non-equilibrium state. Generated coherence is then transferred to energy at the end of the stroke leading to the desired thermal state. In the long cycle time limit, the state follows the adiabatic change in the Hamiltonian  and the cycle converges to the reversible Carnot cycle.

The switching points between strokes are defined in terms of the bath temperatures $T_c,\,T_h$ and oscillator frequencies $\omega_1,\omega_2,\omega_3$ and $\omega_4$. Defining the compression ratio ${\cal{C}}=\omega_{max}/\omega_{min}=\omega_1/\omega_3$ , the cycle is completely determined by $\omega_{min} = \omega_3$, $\cal C$ and bath temperatures. 

At the four corners the WM is in a thermal state: $\hat{\rho}_{th}\b{\omega,T}\equiv Z^{-1}\exp \b{-
\hat{H}\b{\omega}/k_B T}$, where $Z$ is the partition function (see Fig. \ref{fig:H_vs_omega} Panel (a)). The adiabatic strokes conserve the working medium's entropy. Therefore, the populations satisfy  $n_2 = 1/\b{\exp\b{\f{\hbar \omega_2}{ k_B T_h}}-1}= n_3 = 1/\b{\exp\b{\f{\hbar \omega_3}{ k_B T_c}}-1}$ and
 $n_1=n_4$, where $n_i$ is the population at  the $i$'th corner. This leads to the condition
 \begin{equation}
\frac{\omega_3}{\omega_2}=\frac{\omega_4}{\omega_1}=\frac{T_c}{T_h}~~.
\label{eq:omega_carnot_cond}
 \end{equation}
From the cycle definition $\omega_1>\omega_2$ and using Eq. (\ref{eq:omega_carnot_cond}) we obtain a lower bound for the compression ratio 
\begin{equation}
    {\cal C}>\f{T_h}{T_c}~~.
    \label{eq:compression}
\end{equation}

The Carnot-shortcut obtains non-vanishing power by utilizing Shortcut To Equillibration (STE)  protocols \cite{dann2018shortcut} during the open-expansion and open-compression strokes (see \ref{ap:STE}), and Shortcuts To Adiabaticity (STA) protocols for the adiabatic expansion and compression (see  \ref{ap:STA}). 

Shortcut to equilibrium protocols are designed to manipulate an open quantum system between thermal states. The construction of the these protocols is based on  the inertial theorem \cite{dann2018inertial} and the NAME \cite{dann2018time}. 
Starting from a stationary Gibbs state the driving is first accelerated (complying with the inertial theorem) and then decelerated, leading to the target Gibbs state at the end of the stroke.
In the present open-expansion process we modify the frequency from $\omega_1$ to $\omega_2$ while the system interacts with a bath at temperature $T_h$ (or in the open-compression from $\omega_3$ to $\omega_4$ at bath temperature $T_c$). This protocol balances coherence generation and dissipation to achieve a Gibbs state, diagonal in the energy representation. 

The protocol duration can be varied within the framework of the inertial approximation with negligible deviations from optimal fidelity \cite{dann2018shortcut}. Once the protocol duration is cut short, the STE requires a rapid change of the Hamiltonian, generating more intermediate coherence. This leads to larger dissipation, resulting in an increase of the entropy production and work cost.

On the adiabats,  frictionless protocols  are constructed employing  the Lewis-Riesenfeld invariant \cite{lewis1969exact,chen2010fast,chen2011lewis}.  These protocols vary the oscillator frequency non-adiabatically, generating coherence at intermediate times and storing energy in the WM.  At the end of the stroke, the coherence is fully extracted, leading to a total zero work cost. Since all stored energy is retrieved when the protocol is completed, we consider these processes as an analog to catalysis \cite{kosloff2017quantum}.
In principle, these protocols can be achieved almost instantaneously. However, this implies a temporary storage of an infinite amount of energy in the WM \cite{chen2010fast,chen2011lewis,stefanatos2017minimum}. To comply with practical physical considerations,  we choose a constant stroke duration for the adiabats, which is consistent with the inertial theorem. Moreover, the cycle-time is dominated by the open-strokes, therefore the time allocated to the adiabats does not alter the performance qualitatively.  A more complete description accounts for the power dissipated in the controller \cite{torrontegui2017energy,tobalina2018energy}, this contribution is out of the scope of the present analysis.

Control of the cycle time allows a transition between two extreme operation modes. Short cycle times produces finite power ${\cal{P}}=-W/\tau_{cycle}$, on account of a reduction in efficiency, due to quantum friction associated with fast driving. The non-adiabatic driving generates coherence, which is dissipated to the bath. As a result, the useful work decreases, leading to a decrease in efficiency. In contrast, the long cycle time regime optimizes the efficiency while leading to a vanishing power output. Such cycle performance demonstrates the trade-off between power and efficiency. 

Correspondence between the Carnot-shortcut cycle and the ideal classical result is obtained in the quantum adiabatic limit (diverging stroke times). In this limit,
the WM state remains on the energy shell along the whole cycle 
and the STE and STA strokes converge to reversible isothermals and adiabats. The optimal work extraction becomes 
\begin{eqnarray}
\begin{array}{lll}
    W_{C} =& \hbar\Delta\omega_{32}\b{n_{2}+1}+\hbar\Delta \omega _{14}\b{n_{1}+1} 
    +k_{B}\b{T_{h}-T_{c}}\rm{ln}\b{\f{n_{1}}{n_{2}}}~~,
    \end{array}
\label{eq:Work_ideal}
\end{eqnarray}
where $\Delta \omega_{ij} =\omega_i-\omega_j$. In the high temperature limit Eq. (\ref{eq:Work_ideal}) obtains the elegant form
\begin{equation}
W_C=-k_B\, T_h\eta_C\rm{ln}\b{{\cal C}}~~.
    \label{eq: asymp work}
\end{equation}

 \subsection{Performance of the Carnot-shortcut cycle}
\label{subsec: SC results}

\begin{figure}[htb!]
\centering
\includegraphics[width=7cm]{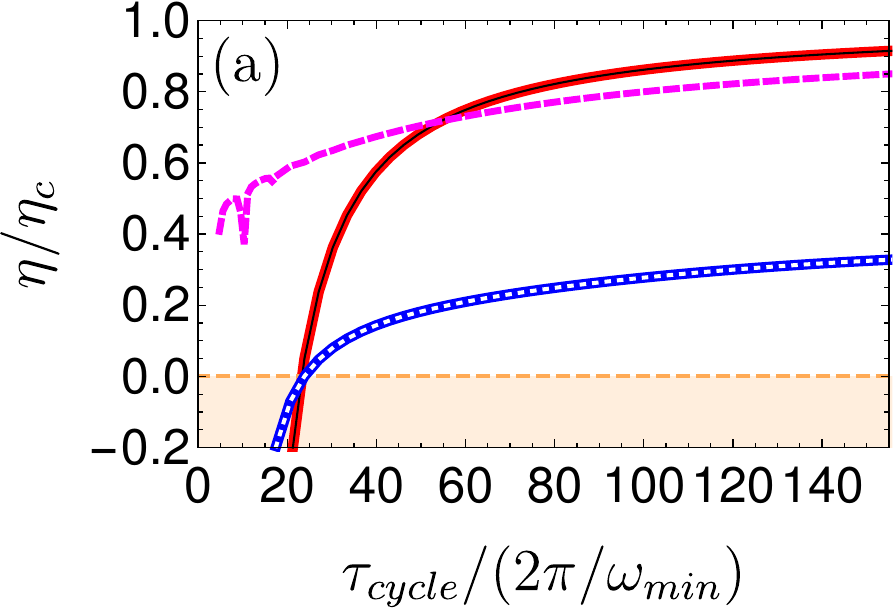}
\includegraphics[width=7cm]{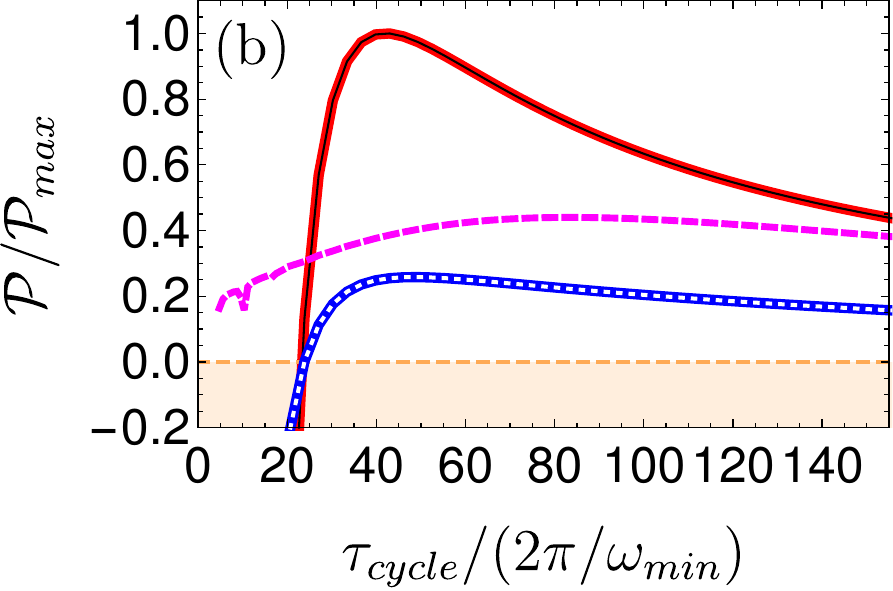}
\caption{(a) Efficiency and (b) power as a function of cycle-time $\tau_{cycle}$. The light orange background indicates negative output power;
a dissipator.
Here, $\eta_C=0.375$, ${\cal{P}}_{max}=6.3\cdot10^{-3}\,\rm{a.u}$.  When $\tau_{cycle}>\tau_{trans}$ the Carnot-analog cycle operates as an engine, for $\tau_{cycle}<\tau_{trans}$ the cycle operates as a dissipator.}
\label{fig:effpowervstau}
\end{figure}

 The performance of the  Carnot-shortcut cycle is analysed by varying the duration of the open strokes, while keeping the time allocation of the  adiabatic strokes fixed. We find two characteristic operational modes: Engine and dissipator. These are associated with different driving speed regimes. For `slow' to `moderate' driving speeds (`long' or 'medium' cycle-times) the cycle operates as an engine (negative work output and positive efficiency, using the convention that outgoing energy is negative). For decreasing cycle-times the work output of the {\it{open-expansion}} decreases (in absolute value) and the work required to perform the  {\it{open-compression}} stroke increases. This leads to a reduced efficiency, see Fig. \ref{fig:effpowervstau} Panel (a). Once the cycle-time is reduced below $\tau_{trans}=23.87\, \b{2\pi/\omega_{min}}$, where $\omega_{min}=5\, \rm{a.u}$, the cycle operates as a dissipator, converting net positive work to heat, which is in turn dissipated to the cold bath. This leads to negative efficiency for $\tau_{cycle}<\tau_{trans}$, as shown in Fig. \ref{fig:effpowervstau} Panel (a). 
 
 \begin{figure}[htb!]
\centering
\includegraphics[width=7cm]{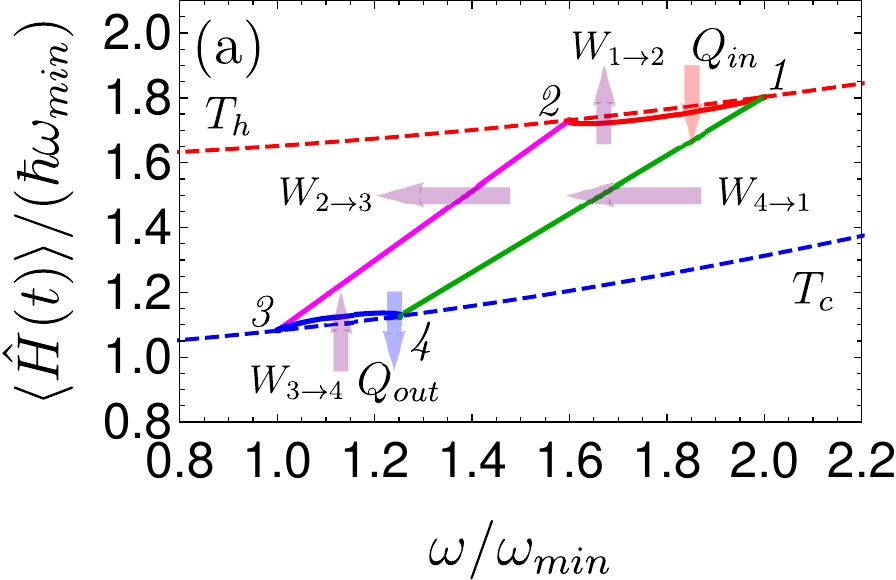}
\includegraphics[width=7cm]{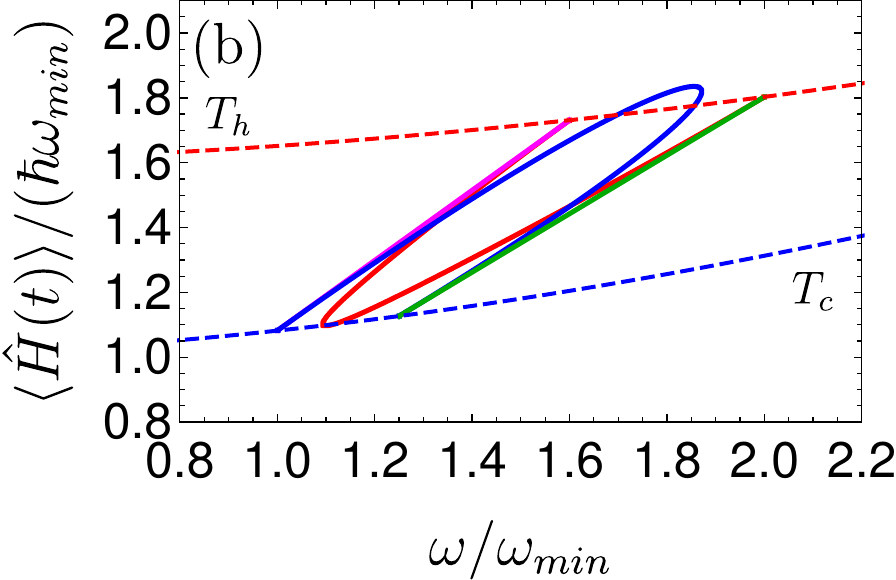}
\includegraphics[width=7cm]{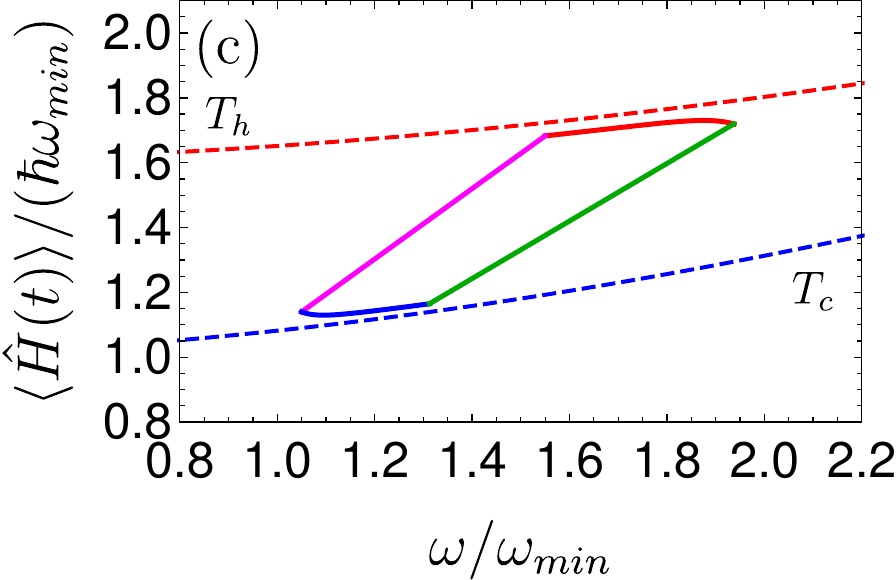}
\includegraphics[width=7cm]{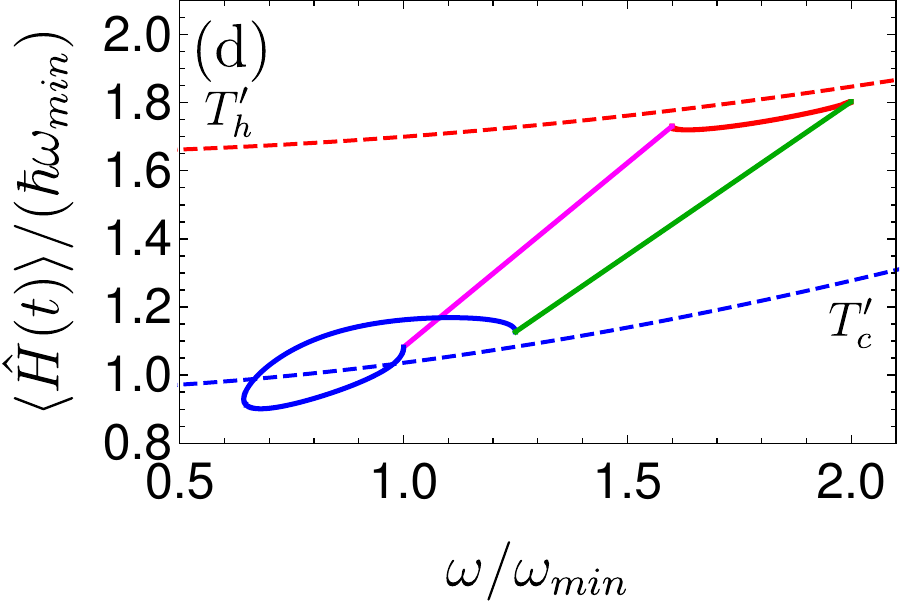}\\
\caption{A map the Carnot-analog cycles: Energy as a function of the oscillator frequency $\omega$. The Carnot-shortcut cycle for cycle-times: (a) $\tau_{cycle}=250$, (b) $\tau_{cycle}=17.5$, the (c) Endo-global cycle and (d) Endo-shortcut cycle for $\tau_{cycle}=250$ (units of $2\pi/\omega_{min}$). Hot (red) and cold (blue) isohtherms are indicated by dashed lines, in panel (d) the isotherms are displaced relative to the previous panels. Cycle strokes are color coded by: {\it{open-expansion}}-red, {\it{adiabatic expansion}}-purple, {\it{open-compression}}-blue and  {\it{adiabatic compression}}-green. Model parameters are presented in table \ref{table:cycle parameters}}
\label{fig:H_vs_omega}
\end{figure}
The transition between dissipator and engine operational modes is caused by the large energy dissipation, which accompanies fast driving. Fig. \ref{fig:H_vs_omega} displays a map of the cycle in the energy frequency plane. For slow driving, Panel (a), the open-strokes follow the isotherms, leading to a Carnot-analog engine that obtains close to optimal efficiency $\eta_C=0.375$. When the cycle-time shortens, the energy deviates significantly from the isotherms, Panel (b), and the cycle operates as a dissipator. Decrease in efficiency is attributed to fast driving, that requires large generation of coherence. Consequently, dissipation is increased, eventually canceling the useful extracted work. Geometrically, the reduction in work can be witnessed as a reduction in the area confined by the cycle, in the $\mean{\hat{H}},\omega$ plane. 
 
Since the adiabats are preformed by frictionless protocols, the cycle's operational mode is dictated by the open strokes. In these strokes, decreasing the stroke times costs work, eventually damaging the performance. Such an effect is naturally described in the simplified asymptotic large cycle time limit. Asymptotically, the work cost scales as $1/t$ \cite{dann2018shortcut} and the power behaves as, (Cf. \ref{sec:friction action})
\begin{equation}
    {\cal{P}}-\f{|W_{C}|}{\tau_{cycle}}\propto -\f{F}{\tau_{cycle}^2}~~,
    \label{eq:P eq prop}
\end{equation}
where $W_{C}$ is the ideal work, Eq. (\ref{eq:Work_ideal}), and $F$ is the friction action (see \ref{sec:friction action}). The three terms of Eq. (\ref{eq:P eq prop}) are positive for an engine operation, inferring that there exists a maximum in power for a finite cycle-time $\tau_{cycle}$ \footnote{Eq. (\ref{eq:P eq prop}) can be written as a second order polynomial in the parameter $x=1/\tau_{cycle}$, with $d^2{\cal{P}}/dx^2<0$. }. Furthermore, for sufficiently small $\tau_{cycle}$ Eq. (\ref{eq:P eq prop}) implies that  the power output is negative and the cycle becomes a dissipator. This leads to the approximate relation between the cycle-time at maximum power $\tau^*$ and the transition time $\tau_{trans}$: $\tau^*=2 \tau_{trans}$. Fig. \ref{fig:effpowervstau} demonstrates this approximate relation.
Overall, 
reducing the cycle-time increases the invested work during the {\it{open-compression}} stroke, and reduces the work output in the {\it{open-expansion}}, in accordance with the thermodynamic principles.

For increased cycle-times, the efficiency of the cycle improves, obtaining the Carnot bound asymptotically (see Fig. \ref{fig:effpowervstau} Panel (a)). Optimal power is achieved for relative short cycle-times $\tau_{max{\cal{P}}}=43 \b{2\pi/\omega_{min}}$, (see Fig. \ref{fig:effpowervstau} Panel (b)). We have also analyzed the performance at elevated temperatures, maintaining a constant temperature ratio. Asymptotically, the power increased linearly, Eq. (\ref{eq: asymp work}), and the efficiency at maximum power converged to the value $\eta_{max{\cal P}}^{*}$ slightly larger than $\eta_{CA}$. This implies that this cycle does not operate in the weak dissipation limit \cite{esposito2010efficiency}.
\begin{figure}[htb!]
\centering
\includegraphics[width=8cm]{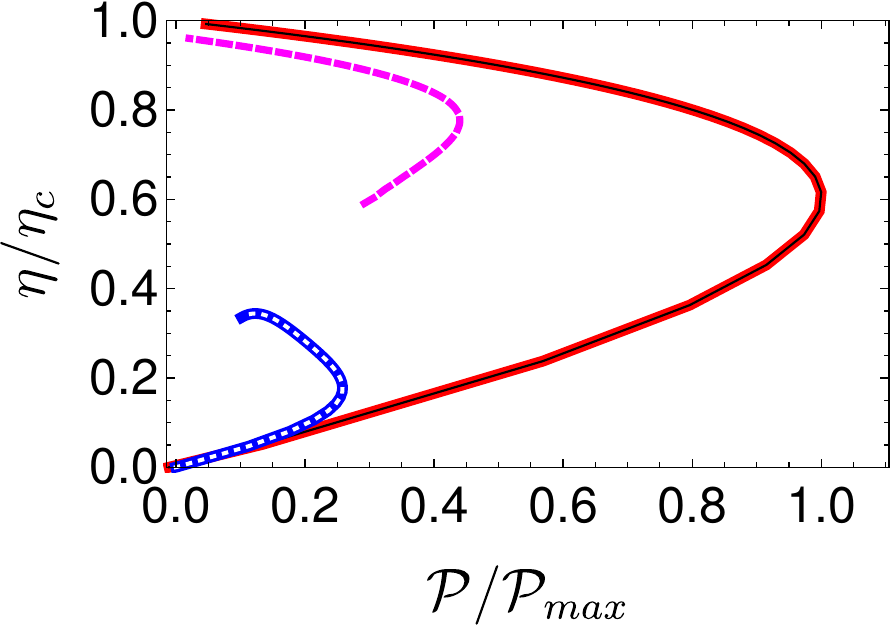}
\caption{Normalized efficiency as a function of the normalized power for the  Carnot-shortcut (red, with thin black line), Endo-global (thin dashed magenta) and Endo-shortcut (blue with white dots) cycles . The Carnot efficiency is $\eta_C=0.375$, and the value of the efficiency at  maximum power $\eta_{max{\cal P}}=0.62\eta_C$ slightly greater than the Curzon-Ahlborn efficiency  $\eta_{CA}=0.56\eta_C$.}
\label{fig:effvspower}
\end{figure}

\section{Quantum signature: Endoreversible Shortcut and Global Carnot cycles}
\label{sec: Endoreversible}

 Endoreversible cycles are defined by four corners for which the WM is in a Gibbs state with temperatures $T\neq T_{bath}'$. As a result, finite heat flow between the engine and bath occurs and the WM is never in equilibrium with the heat bath. In the following, primes designate the bath temperatures of the endoreversible cycle.

The Endo-shortcut cycle is constructed in a similar fashion to the  Carnot-shortcut cycle Sec. \ref{sec: Carnot shortcut cycle}, including two STE and two STA protocols.  For such a cycle, the STE protocols are modified to transform the state between non-equilibrium Gibbs states, Cf. \ref{ap:non-thermal STE}. As a result,  the four corners are maintained in a non-equilibrium Gibbs state $\hat{\rho}\b{\omega,T} = Z^{-1}\exp \b{-\hat{H}\b{\omega}/k_B T}$, keeping the same frequencies ($\omega_1-\omega_4$), where $T$ is the working medium's internal temperature which differs from the bath temperatures $T_c'$ and $T_h'$. The bath temperatures of the Endo-shortcut cycle satisfy $T_c'<T_c$ and $T_h'>T_h$, relative to the Carnot-shortcut bath temperatures $T_c$ and $T_h$, see Table \ref{table:cycle parameters} for a summary of the various cycle parameters. During the open-expansion the STE transfers the oscillator from $\hat{\rho}\b{\omega_1,T_h}$ to $\hat{\rho}\b{\omega_2,T_h}$, and similarly in the open-compression ($\omega_4\ra \omega_1$ with $T=T_c$), see Fig. \ref{fig:H_vs_omega} Panel (d).

Alternatively, the  {\it{Endo-global}} cycle operates as a Carnot-analog cycle where the strokes are performed with a pre-defined constant adiabatic speed $|\mu|=|\dot{\omega}/\omega^2|=\rm{const}$.
The strokes are defined by four frequencies $\omega_1^g, \omega_2^g,\omega_3^g,\omega_4^g$, and $\mu$. These determine the stroke duration $t_f$ and protocols $\omega\b t=\omega_i/\b{1-\mu \omega_i t}$ for a stroke starting at $\omega_i$ and ending at $\omega\b{t_f}$, Cf.  \ref{apsec:Endo-global cycle}. 

 Propagators of the free dynamics, Eq. (\ref{eq:Ham}), can be obtained explicitly in terms of an operator basis in  Liouville space Cf.  \ref{apsec:Endo-global cycle}. These propagators construct the WM unitary transformations of the adiabats. Next, the same free dynamics solution is used to derive the NAME, employed to generate the propagator of the  {\it{open-expansion}}, ${\cal{U}}_h$, and {\it{open-compression}}, ${\cal{U}}_c$, strokes \cite{dann2018time}. Combining the four strokes forms the Global cycle.

 \begin{figure}[htb!]
\centering
\includegraphics[width=8cm]{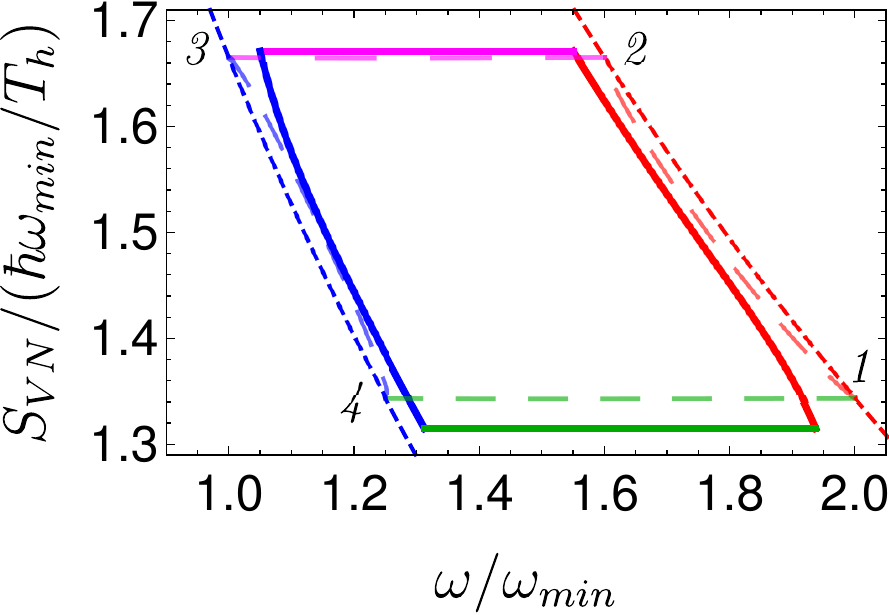}
\caption{ Cycle map: von Neumann Entropy $S_{v.n}$ as a function of the oscillator frequency $\omega\b t$ for the Endo-global cycle (thick continuous lines) and Carnot-shortcut cycle (transparent long dashed), in the slow driving regime $\tau_{cycle}= 250 \b{2\pi/\omega_{min}}$. The Endoreversible cycle operates between the hot (short-dashed red) and cold (short-dashed blue) isotherm lines of temperatures $T_h=8$ and $T_c=5$.}
\label{fig:S_vs_omega_global}
\end{figure}

In the following, we analyze an Endo-global Carnot cycle operating between baths of temperatures $T_c$ and $T_h$, with frequencies $\omega_1^{g}= T_h^{g}\b{\omega_1/T_h}$, $\omega_3^{g} = T_c^{g}\b{\omega_3/T_c}$, $\omega_2^{g}=\omega_3^{g}\b{T_h^{g}/T_c^{g}}$ and $\omega_4 = \omega_1^{g} (T_c^{g}/T_h^{g})$, where $T_c^{g}=5.25$ and $T_h^{g}=7.75$. The frequencies are chosen so as to comply with an ideal Carnot cycle (or  Carnot-shortcut cycle) operating between temperatures $T_c^{g}$ and $T_h^{g}$, in the long cycle-time limit. In this limit, the state at the four corners becomes isoentropic with the states of the Carnot-shortcut cycle. As a result, the power of the two cycles coalesce at long cycle times, see Fig. \ref{fig:effpowervstau} Panel (b). Thermodynamic analysis is carried out on the limit-cycle \cite{feldmann2004characteristics}, defined by the cycle parameters above.

\subsection{Cycle performance comparison}
\begin{figure}[htb!]
\centering
\includegraphics[width=10cm]{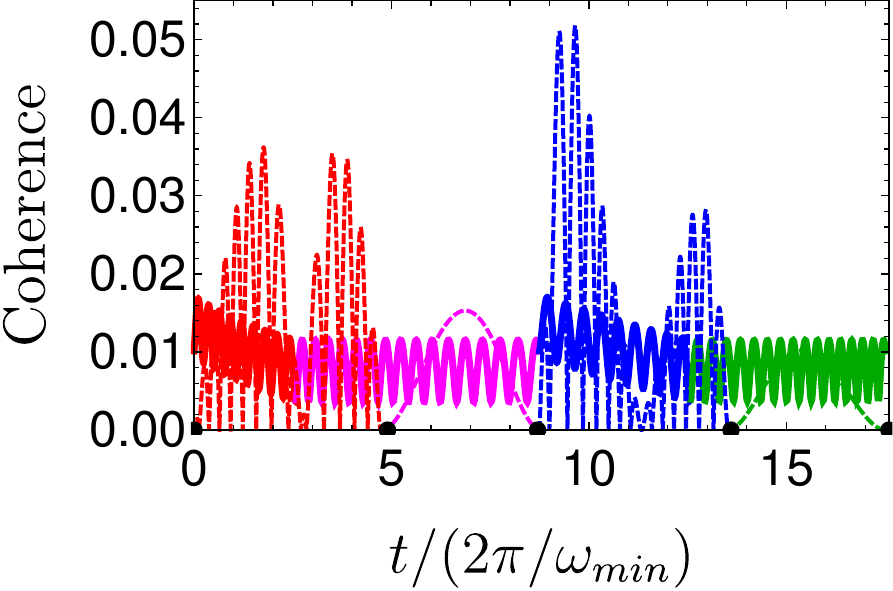}
\\
\caption{Coherence as a function of time for the Endo-global (continuous lines) and Carnot-shortcut (dashed lines) engines for a complete cycle with cycle-time $\tau_{cycle} =8$ (units of $2\pi/\omega_{min}$). Cycle strokes are color coded as in Fig. \ref{fig:H_vs_omega}. Endo-global engine is characterized by a global coherence operation, maintaining a non-vanishing coherence throughout the whole cycle. In contrast, the Carnot-shortcut exhibits branch coherence operation, coherence is created `locally' during each stroke, with initial and final diagonal Gibbs states. Note, that the coherence of the Carnot-shortcut cycle vanishes at the intersections between adjacent strokes. The Endo-shortcut shows a similar behaviour as the Carnot-shortcut, it has been removed for clarity. }
\label{fig:cohvstime}
\end{figure}

Performance of the three cycles is compared with an emphasis on the role of coherence.
 The cycle output is determined by the operational mode and the cycle parameters:  Cycle frequencies, bath temperatures and cycle-time, Cf. Table \ref{table:cycle parameters}. The Shortcut cycles (Carnot and endoreversible) are characterized by a branch coherence operation, where coherence is created `locally' during each stroke, with initial and final diagonal Gibbs states. This can be seen in Fig. \ref{fig:cohvstime}, where the dashed lines represent the coherence of the Carnot shortcut cycle which vanish after each stroke. Conversely, in the limit-cycle of the Endo-global  engine, coherence is maintained throughout the cycle (globally) (non-vanishing continuous line in Fig. \ref{fig:cohvstime}). As a result, coherence generated in one stroke can be utilized along the subsequent strokes. 
  
 This comparison is unbiased, since the same bath temperatures are considered for the Carnot-shortcut and Endo-global cycles. In the Endo-shortcut cycle, despite of the larger  temperature difference there is no foreseen advantage.  
 

The Carnot-shortcut cycle shows a superior performance at moderate and long cycle-times, Figs. \ref{fig:effpowervstau} and \ref{fig:effvspower}. In this operational regime, the efficiency, power, maximum power, and efficiency at maximum power exceed both the Endo-global and Endo-shortcut cycles. However, for short cycle-times, both shortcut cycles become dissipators, producing negative power, while the Endo-global cycle continues to operate as an engine. 

 \begin{figure}[htb!]
\centering
\includegraphics[width=8cm]{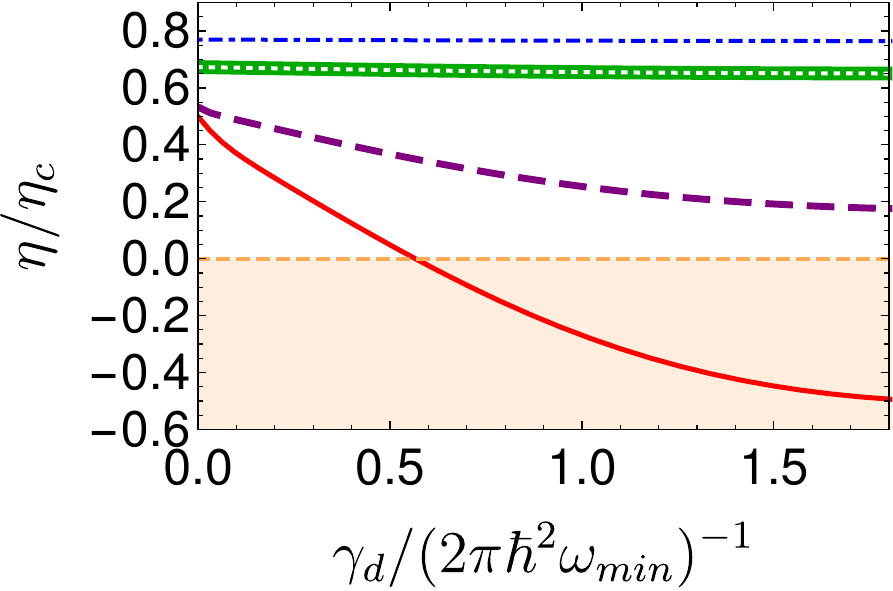}
\\
\caption{Normalized efficiency as a function of the dephasing strength $\gamma_d$, for  $\tau_{cycle} =8$ (red continuous line), $\tau_{cycle} =12$ (purple-dashed), $\tau_{cycle} =40$ (thick green with white dots), and $\tau_{cycle} =80$ (blue-dot-dashed) (units of $2\pi/\omega_{min}$). For short cycle-times, coherence becomes central to the engine's operation. In this regime, increasing the dephasing strength leads to a reduction of efficiency. On the other hand, when the driving is slow, generation of coherence is small and the dephasing does not affect the engine's performance. Note, that for negative efficiency the cycle operates as a dissipator.}
\label{fig:effvskdephasing}
\end{figure}

This behaviour compels us to reexamine the question: When is a quantum heat engine quantum?
An intuitive approach defining a quantum signature, is that by observing the engine operation it will cease to produce useful output. To be more precise we can imagine continuously monitoring the energy of the WM on the adiabatic strokes. 
As a result, the state will relax to a diagonal state in the energy representation and coherence
vanishes. The back action of the weak measurement is therefore equivalent to pure dephasing \cite{silberfarb2005quantum,konrad2010monitoring}, 
described by the master equation:
\begin{eqnarray}
\begin{array}{ll}
   \f{d}{dt}\hat{\rho}\b t=&-\f{i}{\hbar}\sb{\hat{H}\b t,\hat{\rho}\b t}-\gamma_d \sb{\hat{H}\b t,\sb{\hat{H}\b t,\hat{\rho}\b t}}~~,
   \end{array}
   \label{eq: deph term}
\end{eqnarray} 
where the parameter $\gamma_d$ scales the dephasing rate. The explicit derivation
is found in \ref{sec:dephasing}. The influence of the dephasing rate on the engine efficiency is shown in Fig. \ref{fig:effvskdephasing}. For long cycle-times 
where the nonadiabatic effects are small the influence of dephasing is minor.  
Conversely, for short cycle-times the engine operation requires coherence and the 
strong dephasing nulls the power output. Similar behaviour has been observed in the sudden Otto cycle and the two-stroke NV engines \cite{kosloff2017quantum,feldmann2016transitions,klatzow2019experimental}.

Another viewpoint on quantum signatures arises from a comparison with the classical counterpart. 
Introducing strong dephasing effectively leads to a pure stochastic mode of operation, where the system is solely characterized by its population in the energy representation and the energy levels. The improved efficiency, in the small cycle time regime, demonstrates that for fast driving the coherent operation mode outperforms the stochastic work extraction mechanism. This highlights the quantum advantage in this regime.  The relative increase of efficiency under coherent operation also constitutes a thermodynamic measurable quantity that indicates the existence of a pure quantum phenomena. Therefore, it is considered as a quantum signature. In essence, coherence emerges when the WM state is in a superposition of eigenstates of the instantaneous Hamiltonian, implying that the superposition nature is the origin of the increased efficiency and quantum signature.



\subsection{Geometric representation of coherence}

\begin{figure}[htb!]
\centering
\includegraphics[width=6cm]{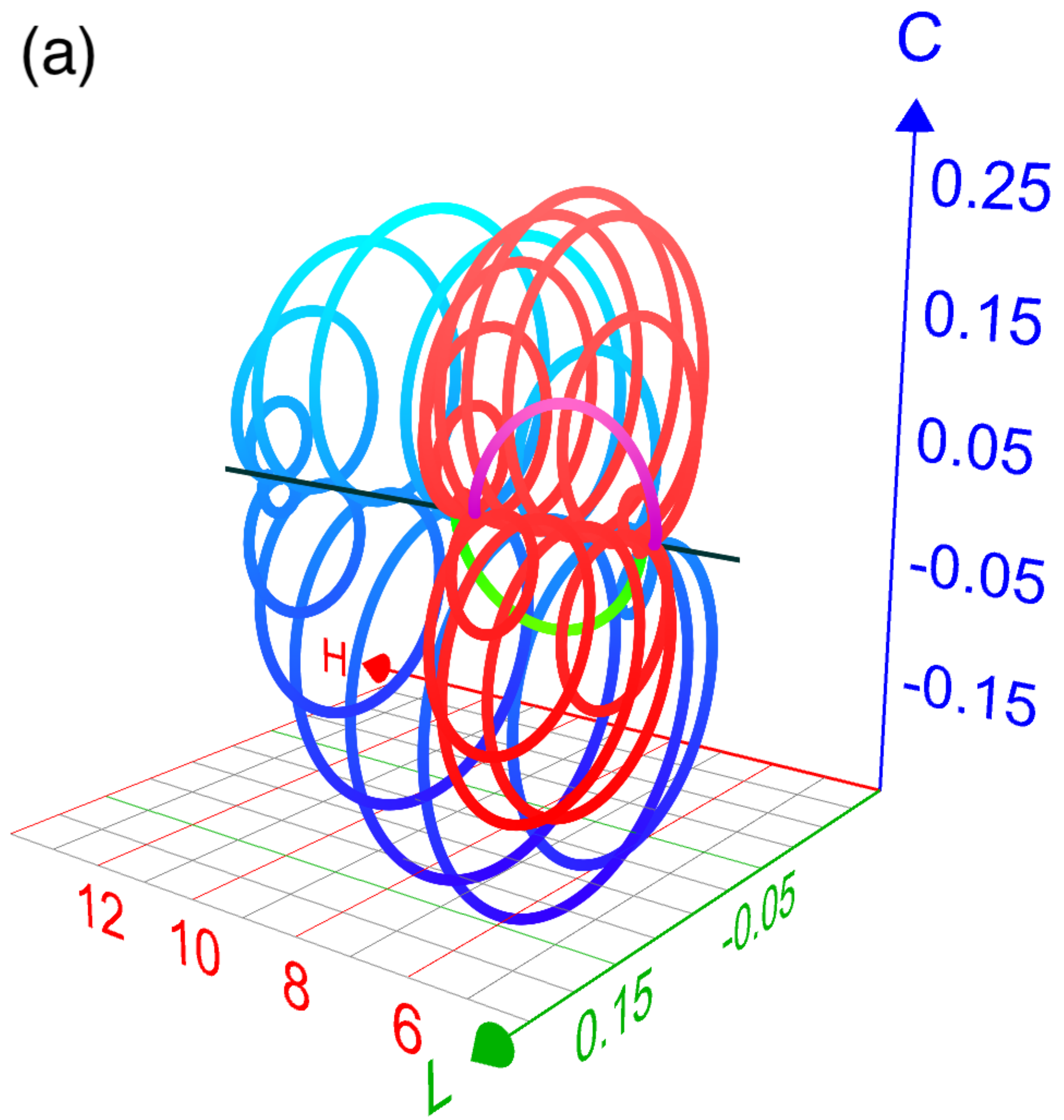}
\includegraphics[width=6cm]{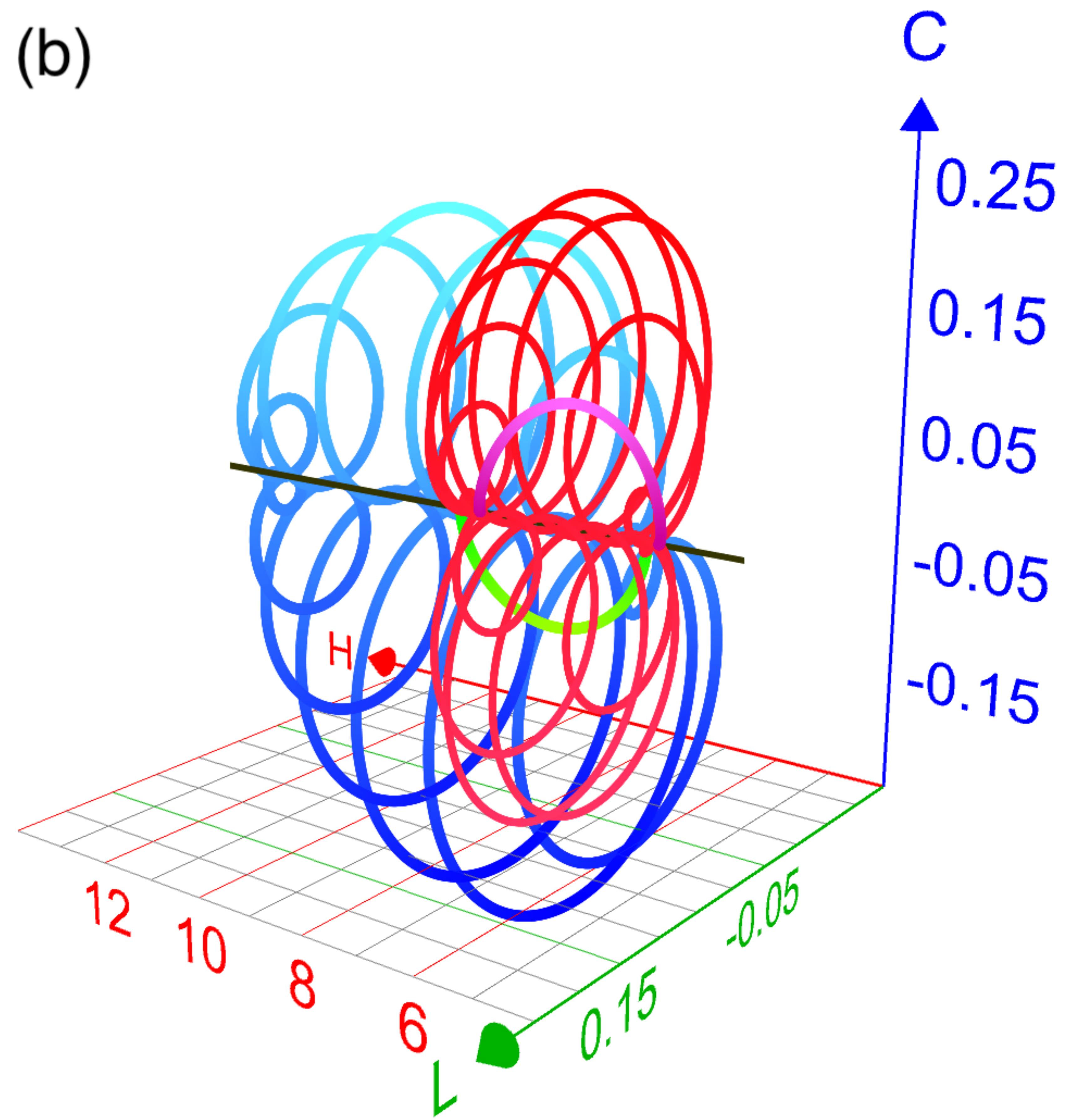}
\includegraphics[width=6cm]{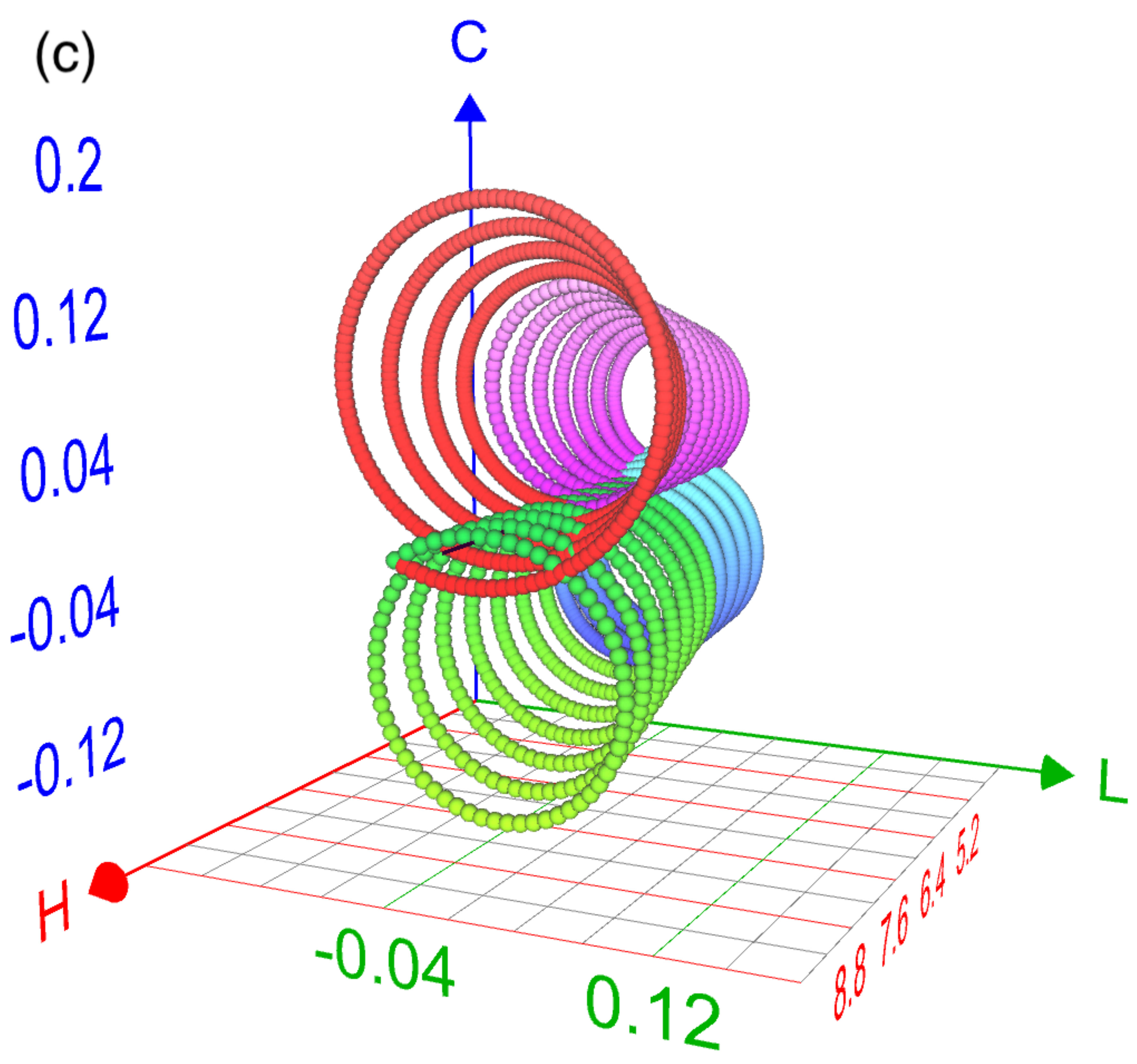}
\includegraphics[width=6cm]{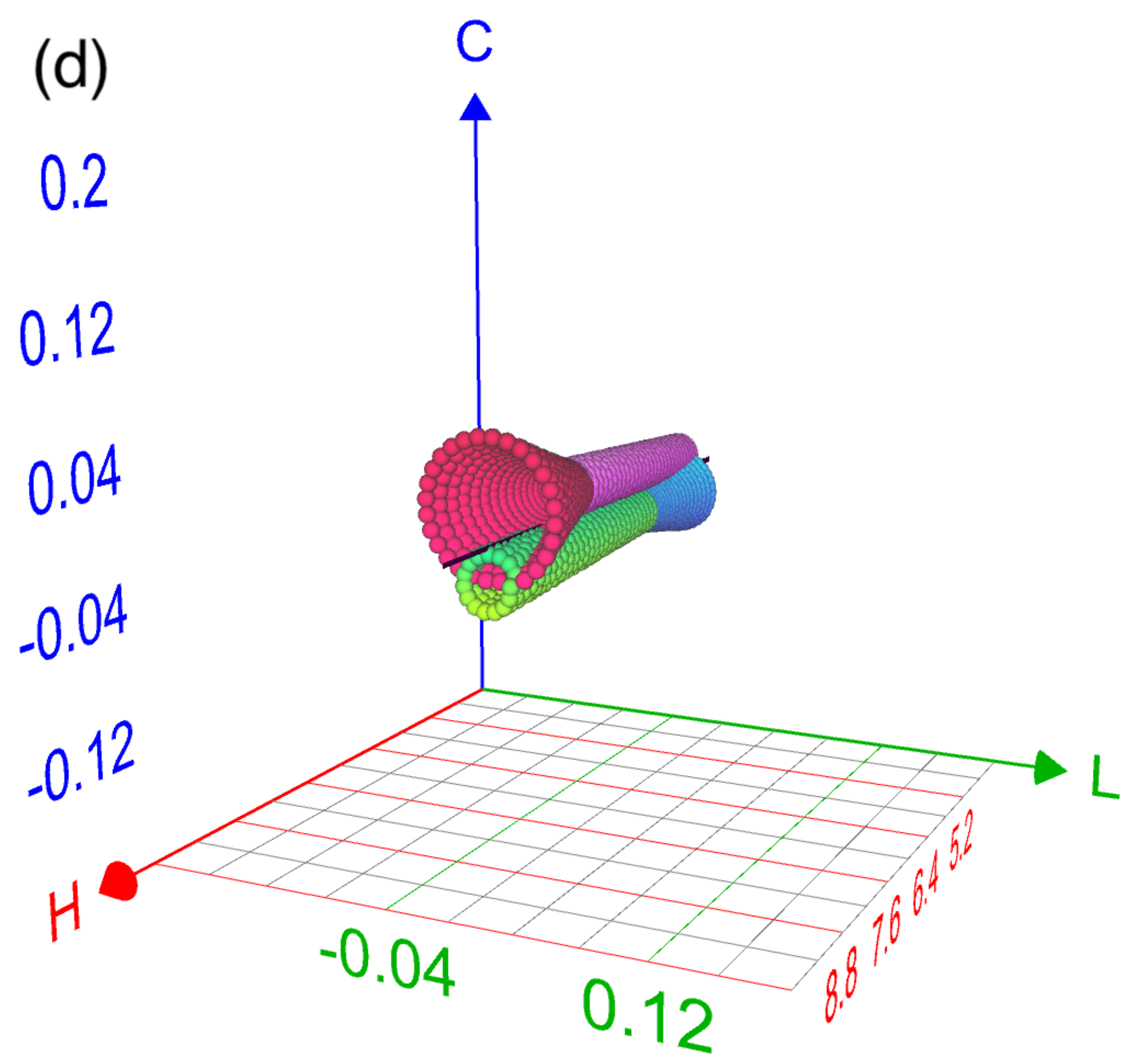}
\\
\caption{Cycle trajectory in the $\{ \mean{\hat{H}},  \mean{\hat{L}}, \mean{\hat{C}}\}$ observable vector space, where $\hat{L}$ is the Lagrangian, $\hat{C}$ is the position momentum correlation operator and $\hat{H}$ is the Hamiltonian Eq. (\ref{eq:Ham}) \cite{kosloff2017quantum}. The red and blue lines display the open-expansion and compression strokes and the purple and green lines display the expansion and compression unitary strokes. Panel (a) shows the Shortcut cycle (b) Endo-shortcut cycle for a fast driving $\tau_{cycle}=17.5$ (c),(d) Endo-global cycle for  $\tau_{cycle}=8$ and $\tau_{cycle}=32$ (units of $2\pi/\omega_{min}$).
The black line designates zero coherence $\mean{\hat{L}}=\mean{\hat{C}}=0$. Note, that in the shortcut trajectories coherence vanishes at the cycle's four corners.  }
\label{fig:3D}
\end{figure}

To illuminate the interplay between coherence and energy throughout the Carnot analog cycles, we present a geometrical representation of the WM state.  
Generally, a Gaussian state of the Harmonic oscillator is fully characterized by a basis of four operators forming a closed Lie algebra. Moreover, assuming an initial Gaussian form, the WM state is canonical invariant under the equations of motion of the various strokes \cite{rezek2006irreversible,kosloff2017quantum}. We chose a time-dependent operators basis  $\{\hat{H}\b t,\hat{L}\b t,\hat{C}\b t\}$,  see Sec. \ref{sec: Dynamics of Carnot-analog} and \ref{apsec:Endo-global cycle}. This leads to the geometrical representation in terms of the vector $\v v \b t= \{\mean{\hat{H}},\mean{\hat{L}},\mean{\hat{C}}\}^T$, shown in Figure \ref{fig:3D}.

Coherence in the energy basis is related to the
operators $\hat{L}\b t$ and $\hat{C}\b t$ which do not commute with $\hat{H}\b t$,   \ref{apsec:Endo-global cycle} Eq.  (\ref{eq:coherence}). The shortcut cycles, shown in Panels (a) and (b) display significant coherence generation, with an overshoot in energy during the open-strokes. The Lewis-Riesenfeld protocols of the adiabatic strokes lead to $\mean{\hat{L}\b t}\approx 0$ \cite{chen2011lewis}. In this cycle, all the connecting corners between strokes are on the line $\mean{\hat{L}}=\mean{\hat{C}}=0$. The trajectory of the Endo-global cycle encircles the zero coherence line $\mean{\hat{L}}=\mean{\hat{C}}=0$, while never coinciding with it. As expected, short cycle-times lead to greater coherence, Panel (c).

\begin{table}
\caption{Cycle parameters}
\label{table:cycle parameters}
\begin{center}
\begin{tabular}{ |p{3.0cm}||p{3.6cm}|p{3.6cm}|p{3.6cm}|  }
 \hline
 Cycle type & Carnot-shortcut & Endo-Shortcut & Endo-global\\
 \hline
 Baths temperatures   & $T_c=5~~~~~~~~~~~$ $T_h=8$  &$T_c'=5.25~~$ $T_h'=7.75$ &    $T_c=5~~~~~\,$ $T_h=8$\\
 \hline
 Cycle frequencies&   $\omega_1=10~~~~~~~~~~~$ $\omega_2=\omega_3\f{T_h}{T_c}=6.25$ $\omega_3=5~~~~~~~~~~~$ $\omega_4=\omega_1 \f{T_c}{T_h}=7.5$   & $\omega_1=10~~~~$ $\omega_2=6.25~~$ $\omega_3=5~~~~~$ $\omega_4=7.5$   &$\omega_1^{g}=9.6875$ $\omega_2^{g}=7.75$ $\omega_3^{g}=5.25$ $\omega_4^{g}=6.5625$\\
 \hline
 Operation type &Branch coherence& Branch coherence&  Global coherence\\
 \hline
\end{tabular}
\end{center}
\end{table}

 \section{Discussion}
 \label{sec:discussion}
 A family of finite power quantum Carnot-analog engines is investigated to determine the role of coherence in the engine operation. In the construction of the models, we have overcome the difficulty of describing thermalization in the presence of non-adiabatic driving \cite{dann2018time}, including an explicit description of the working medium dynamics. This enables a comparison to other finite-time quantum cycle as the Otto cycle \cite{kosloff2017quantum}.

Carnot and Otto engines have been the keystones in the study of quantum heat devices. These engines differ by their thermalization strokes and share their unitary adiabatic strokes.
For a fair comparison, we assume the same working medium and bath temperatures.
The Otto and Carnot cycles have different permissible compression ratio range. The Otto cycle is constrained by $\f{T_h}{T_c}\geq{\cal{C}}>1$, while the compression ratio of the Carnot cycle obeys ${\cal{C}}>\f{T_h}{T_c}$. Optimal work for the Otto cycle is achieved when ${\cal{C}}=\sqrt{\f{T_h}{T_c}}$ and vanishes in limits ${\cal C}=1$ and ${\cal C}=\f{T_h}{T_c}$.
On the other hand, the power of the Carnot-analog has a local optimum, and the work output diverges with ${\cal C}$, Eq. (\ref{eq: asymp work}).  

A major performance measure of engines is the efficiency. Otto and the Carnot cycles have different efficiency characteristics, in particular, their dependence on the compression ratio and bath temperatures.
The Otto efficiency $\eta_O = 1-{\cal C}^{-1}$ converges to $\eta_C$ in the limit of zero work. Moreover, optimal efficiency increases conjointly with the work, achieving $\eta_O=\eta_{CA}$ at the optimum power point in the high temperature limit \cite{rezek2006irreversible}.  
Conversely, for the Carnot-shortcut cycle, efficiency $\eta_C$ is constant as function of compression ratio, while the work monotonically increases with $\cal C$.
For this cycle the power obtains its optimum  at a finite compression ratio. In addition, the efficiency at maximum power is found to be greater than the Curzon-Ahlborn efficiency  $\eta_{max{\cal P}}>\eta_{CA}$. 


Control of coherence is the key to obtain finite-power in quantum engines. Shortcut protocols utilize coherence temporally to accelerate the dynamics and enhance the power. On the adiabatic unitary strokes, shortcuts (STA) can be obtained with no direct cost, Sec. \ref{sec: Dynamics of Carnot-analog} \cite{del2014more,beau2016scaling}. For the thermalization strokes, STE protocols enable swift equilibration. This is achieved by generating significant coherence which increases the dissipation of energy and information to the bath. As a result, such a speedup is accompanied by an additional work cost. These protocol are incorporated in the operation of the Carnot-shortcut and Endo-shortcut. The dissipated potential work in the cycle explains the mechanism leading to the power-efficiency tradeoff, and the existence of a minimum cycle time \cite{rezek2006irreversible,kosloff2014quantum,correa2013performance,hoffmann1997endoreversible}. Similar STE protocols can also be used to accelerate the thermalization in the Otto and Stirling cycle.  In addition, STE protocols can be experimentally realized in engines with harmonic working fluid as well as a spin system \cite{rossnagel2016single,von2019spin,funo2019speeding}.

 
 The Endo-global cycle is characterized by non-vanishing coherence throughout the cycle operation,  Fig. \ref{fig:cohvstime}. This is in contrast to the shortcut cycles, which are diagonal in energy representation at the four corners of the cycle. Global coherence enables engine operation at short cycle-times, where the Endo-Shortcut and Carnot-Shorcut become dissipators. This indicates a quantum signature\cite{klatzow2019experimental}. In the presence of  pure-dephasing, the coherence vanishes, reducing power and efficiency. Eventually, strong dephasing transforms the cycle from an engine to a dissipator, a signature of a pure quantum operational mode, see Fig. \ref{fig:effvskdephasing}.

To conclude, the motto of this study follows Carnot's tradition of learning by example. The analysis of a solvable cycle has unraveled the role of coherence. On one hand, coherence is responsible for quantum friction, and on the other hand it allows rapid thermalization and a coherent mode of operation.

 \ack
 We thank Peter Salamon for inspiring this study. 
 We also gratefully acknowledge KITP for their hospitality, and the support of the Adams Fellowship Program of the Israel Academy of Sciences and Humanities. Partial funding was supplied by the National Science Foundation under Grant No. NSF PHY-1748958 and the Israel Science Foundation Grant No.  2244/14. 
 
\appendix

\section{Shortcut to equilibration protocol}
\label{ap:STE}
The STE protocol rapidly transfers the working medium between two Gibbs states with different entropies.   Construction of the protocol relies on the Non-Adiabatic Master Equation (NAME) \cite{dann2018time} and the Inertial Theorem \cite{dann2018inertial}. Derived from first principles, the NAME describes the reduced dynamics of a non-adiabatically driven open quantum system in the weak coupling Markovian regime. The NAME incorporates, as a limit, both the adiabatic and periodic Floquet master equations.

The derivation of the NAME requires  an explicit solution of the closed system dynamics. Utilizing the inertial theorem, we obtain the free propagator $\hat{U}_S \b t$ for slowly `accelerated' external control. For this study, we considered a harmonic oscillator working medium, where the control is achieved by varying the potential. In this case, the inertial condition is associated with a slow change in the adiabatic parameter $\mu=\dot{\omega}/\omega^2$. The explicit solution of free dynamics propagator allows to transform the system-bath coupling to the interaction representation. Using the Born-Markov approximation we derive the generator of the open system dynamics. For this case, we consider an Ohmic Boson bath with Markovian properties. The derivation leads to the reduced system dynamics in the interaction representation 
\begin{eqnarray}
\begin{array}{ll}
\f {d}{dt}\tilde{\rho}_{S} \b t={\cal L} \tilde{\rho}=&k_{\downarrow} \b t\b{\hat{b}\tilde{\rho}_{S}\b t\hat{b}^{\dagger}-\f 12\{\hat{b}^{\dagger}\hat{b},\tilde{\rho}_{S}\b t\}}\\
&+k_{\uparrow}\b t \b{\hat{b}^{\dagger}\tilde{\rho}_{S}\b t\hat{b}-\f 12\{\hat{b}\hat{b}^{\dagger},\tilde{\rho}_{S} \b t\}}~~~.
\end{array}
\label{eq:NAME}
\end{eqnarray}
Here, the density operator in the interaction picture reads $\tilde{\rho}_S\b{t}=\hat{U}_S\b{t,0}\hat{\rho}_S \b t \hat{U}_S^{\dagger}\b{t,0} \label{eq:int} $.
\begin{equation}
k_{\downarrow}\b t=k_{\uparrow} \b t e^{\alpha\b t/k_B T}=\f{\alpha \b t|\v d|^{2}}{4\pi \varepsilon_{0}\kappa \hbar c}\b{1+N\b{\alpha\b t}}~,
\label{eq:k down}
\end{equation}
where $N$ is the occupation number of the Bose-Einstein distribution, $\kappa = \sqrt{4-\mu^2}$, and $\alpha$ is a modified frequency, determined by the non-adiabatic driving protocol \cite{dann2018time}. In terms of the oscillator frequency, the modified frequency is given by
\begin{equation}
\alpha \b t= \sqrt{1-\b{{\dot{\omega}\b t}/\b{2\omega^2 \b t}}^2}\,\, \omega \b t~~.    
\label{eq:alpha1}
\end{equation}
The Lindblad jump operators become $\hat{b}\equiv  \hat{b}\equiv \hat{b} \b 0 =\sqrt{\f{m\omega\b 0}{\kappa\hbar}}\f{\kappa+i\mu}2\b{\hat{Q}\b{0}+\f{\mu+i\kappa}{2m\omega\b 0}\hat{P}\b{0}}$.

In the adiabatic driving limit, $\mu\ra 0$, the Lindblad jump operators converge to the adiabatic creation and annihilation operators  $\hat{b}^{\dagger},\hat{b}\ra \hat{a}^{\dagger},\hat{a}$. Therefore, in this limit, Eq. (\ref{eq:NAME}) reduces to the adiabatic master equation.

The completely positive map generated by Eq. (\ref{eq:NAME}) preserves the Gaussian form. This property is termed canonical invariance \cite{alhassid1978connection,andersen1964exact,rezek2006irreversible,kosloff2017quantum}. Formally, the Gaussian state can also be expressed in a product form 
\begin{equation}
  \tilde{\rho}_S \b t= Z^{-1}e^{\gamma\b t \hat{b}^2}e^{\beta\b t \hat{b}^{\dagger}\hat{b}}e^{\gamma^{*}\b t \hat{b}^{\dagger 2}}~~,  
  \label{eq:Gibbs state}
\end{equation}
where $Z={\tr}\b{e^{\gamma\b t \hat{b}^2}e^{\beta\b t \hat{b}^{\dagger}\hat{b}}e^{\gamma^{*}\b t \hat{b}^{\dagger 2}}}$, the operators $\hat{b}^\dagger \hat{b},\,\hat{b}^2, \hat{b}^{\dagger2}$ vary with $\mu$. Parameters $\gamma$ and $\beta$ are time-dependent functions.

For an initial thermal state, the construction is simplified and the reduced system remains in the following form throughout the entire evolution
\begin{equation}
    \tilde{\rho}_S \b{\beta \b t,\mu \b t}=Z^{-1} e^{\beta \hat{b}^{\dagger }\hat{b} \b \mu}~~,
    \label{eq:rho_7}
\end{equation}
with initial conditions $\beta \b 0=-\f{\hbar \omega \b 0}{k_B T}$ and $\mu\b 0 =0$. Substituting Eq. (\ref{eq:rho_7}) into Eq. (\ref{eq:NAME}) leads to  a non-linear differential equation for $\beta\b t$
\begin{equation}
\dot{\beta}=k_{\downarrow}\b t\b{e^{\beta}-1}+k_{\uparrow}\b t\b{e^{-\beta}-1}~~.
\label{beta_dot}
\end{equation}
Equation (\ref{beta_dot}) constitutes the basis for the STE control scheme. We define $y\b t \equiv e^{\beta\b t}$ and guess a polynomial solution for $y$. This solution should transfer an initial thermal state of frequency 
$\omega\b 0$ to a final thermal state with a frequency $\omega\b{t_f}$, 
which leads to the boundary conditions  
$\beta\b{0}=-\f{\hbar \omega \b 0}{k_B T}$,  $\beta\b{t_f}=-\f{\hbar \omega \b{t_f}}{k_B T}$ and $\mu\b 0 =\mu\b{t_f}=0$. Furthermore, the condition on $\mu$ implies that the state and protocol are stationary at initial and final times, leading to $\dot{\beta}\b{0}=\dot{\beta}\b{t_f}=\ddot{\beta}\b{0}=\ddot{\beta}\b{t_f}$. 

A fifth-degree polynomial is sufficient to comply with all the boundary conditions.  The solution reads
\begin{equation}
y\b{s}=y\b 0 + c_3 \b{t/t_f}^3 + c_4 \b{t/t_f}^4+c_5 \b{t/t_f}^5 ~~,
\label{ap:y_s}
\end{equation}
where $c_3-c_5$ are determined from the boundary conditions $y\b 0=e^{\beta\b 0}$, $y\b{t_f}=e^{\beta \b{t_f}}$, and $\dot{y}\b 0=\dot{y}\b{t_f}=\ddot{y}\b 0=\ddot{y}\b{t_f}=0$.
Next, we insert the solution for $y$, Eq. (\ref{ap:y_s}), into Eq. (\ref{beta_dot}) and solve for $\alpha\b t$. The control $\omega\b t$ is obtained by solving Eq. (\ref{eq:alpha1}) by numerical means.

\subsection{STE for non-thermal initial and final states}
\label{ap:non-thermal STE}
The Endo-shortcut cycle, Sec. \ref{sec: Endoreversible}, includes {\it{open-compression}} and {\it{open-expansion}} strokes between non-equilibrium states. During these strokes, the working medium state is of the Gibbs form
\begin{equation}
    \hat{\rho}\b t = Z^{-1} e^{-\f{\hbar \omega\b 0}{k_B T_i}}~~,
    \label{ap:rho_non_eq}
\end{equation}
where the internal temperatures $T_i$ and $T_f$ differs from the bath temperatures, i.e., $T_i=T_f\equiv T_0\neq T$.

The control protocol $\omega \b t$ for the open-strokes is obtained by a similar reverse-engineering method as in the case of an initial and final equilibrium states ($T_i,T_f=T$). However, since the system-bath interaction leads to non-vanishing decay rates, the initial and final states are non-stationary, which implies $\dot{\beta}\b{0},\dot{\beta}\b{t_f}\neq 0$.
As in the previous construction, we require a continuous change in the control protocol, associated with the restriction $\dot \omega\b{0} =\dot{\omega}\b{t_f} = 0$. Substituting the condition $\dot \omega\b{0} = 0$ into the decay rates of Eq. (\ref{beta_dot}) we obtain the initial value for $\dot{\beta}$ (this is in accordance with the dynamics of the time-independent master equation \cite{breuer2002theory}). This leads to the boundary conditions for $y$: $y\b 0 =y\b 0= \exp \b{-\f{\hbar \omega\b 0}{k_B T_0}}$, $\dot{y}\b 0=\dot{\beta}\b 0\exp\b{-\f{\hbar \omega\b 0}{k_B T_i}}$, $y\b{t_f} = \exp \b{-\f{\hbar \omega\b{t_f}}{k_B T}}$, $\dot{y}\b{t_f}=\dot{\beta}\b{t_f}\exp\b{-\f{\hbar \omega\b{t_f}}{k_B T_0}}$ and $\ddot{y}\b 0=\ddot{y}\b{t_f}=0$. Following a similar derivation as the previous section, we introduce a fifth order polynomial that satisfies the boundary conditions 
to obtain $\omega \b t$.

\section{Adiabatic strokes - Shortcut to adiabaticity protocols utilizing the Lewis Riesenfeld invariant }
\label{ap:STA}
The (thermodynamic) adiabatic strokes are achieved utilizing shortcut to adiabaticity (STA) protocols. These transform a diagonal state in the energy basis, of a frequency $\omega_i$, to a state with the same population, with a final frequency $\omega_f$. These protocols are engineered utilizing the Lewis-Riesenfeld invariant \cite{lewis1969exact}. We follow a similar procedure as presented in Refs. \cite{lewis1969exact,chen2010fast} to construct the STA protocols.

We introduce an Hermitian invariant $\hat{I}$ for the harmonic oscillator algebra, SU(1,1).  Generally, such an invariant can be expressed as sum of the algebra operators
\begin{equation}
    \hat{I}\b t =\f{1}{2}\b{\alpha\b t \hat{Q}^2+\beta\b t \hat{P}^2 +\gamma \b{\hat{Q}\hat{P}+\hat{P}\hat{Q}}}~~,
    \label{ap:I}
\end{equation}
and must satisfy the condition
\begin{equation}
    \f{d\hat{I}}{dt} = \pd{I}{t}+\f{1}{i\hbar}\sb{\hat{I},\hat{H}}=0~~.
    \label{ap:invar_cond}
\end{equation}
To obtain transition-less driving we desire a protocol for which the invariant commutes with the Hamiltonian at initial and final times:
\begin{equation}
    \sb{\hat{H}\b 0,\hat{I}\b 0}=\sb{\hat{H}\b{t_{f}},\hat{I}\b{t_{f}}}=0~~.
    \label{eq:Com I H}
\end{equation}
Since the operators share a common eigenstate basis at these times and the eigenvalues of $\hat{I}$ are stationary, the engineered protocol induces transition between two diagonal states in the energy representation.

 Substituting Eq. (\ref{ap:I}) into Eq. (\ref{ap:invar_cond}) leads to three coupled differential equations for the time-dependent coefficients
 \begin{eqnarray}
 \begin{array}{cc}
\dot{\alpha}= 2 m \omega^2 \gamma  \\
     \dot{\beta} = -\f{2}{m}\gamma \\
     \dot{\gamma} = -\f{1}{m}\alpha+ m\omega^2 \beta
 \end{array}~~.
 \label{ap:coupled_diffs}
 \end{eqnarray}
By defining $\beta \equiv \sigma^2$ and conducting some algebraic manipulations (see Ref \cite{lewis1969exact}) Eqs. (\ref{ap:coupled_diffs}) can be represented in terms of a single differential equation
\begin{equation}
    \f{d}{dt}\b{m^2 \omega^2 \sigma +m^2 \ddot{\sigma}}\sigma+ 3\dot{\sigma}\b{m^2 \omega^2 \sigma +m^2 \ddot{\sigma}}=0~~.
 \label{ap:final dif}
\end{equation}
Solving for  $m^2 \omega^2 \sigma +m^2 \ddot{\sigma}$ and substituting the solution into Eq. (\ref{ap:final dif}) gives
\begin{equation}
    \f{c}{\sigma^2}=m^2 \omega^2 \sigma^2+m^2\sigma \ddot{\sigma}~~,
\label{ap:sigma_dif}
\end{equation}
where, $c$ is an arbitrary real integration constant. Next, we define $\sigma=c^{1/4} \rho$, and substitute the definition into Eq. (\ref{ap:sigma_dif}), to obtain
\begin{equation}
    \hat{I}\b t = \f{1}{2}\b{\f{1}{\rho^2}\hat{Q}^2+\b{m\dot{\rho}\hat{Q}-\rho \hat{P}}}~~,
\end{equation}
with the subsidiary condition (Eq. (\ref{ap:sigma_dif}) 
\begin{equation}
 m^2\omega^2\b t\rho+m^2\ddot{\rho}-\f{1}{\rho^3} = 0~~.
 \label{ap:rho_dif}
\end{equation}
This equation introduces constraints on the protocol $\omega\b t$, which comply with the boundary conditions of $\rho\b t$. The strategy to engineer a transition-less control protocol is to choose $\rho\b t$ such that relations Eq. (\ref{eq:Com I H}) are satisfied. This leads to the following boundary conditions
\begin{eqnarray}
\label{ap:BC}
\begin{array}{ll}
    \dot{\rho}\b 0=\ddot{\rho}\b 0=\dot{\rho}\b{t_{f}}=\ddot{\rho}\b{t_{f}}=0\\
    \rho\b 0=\f 1{\sqrt{m\omega\b 0}}\\
    \rho\b{t_{f}}=\f 1{\sqrt{m\omega\b 0}}\sqrt{\f{\omega_{0}}{\omega_{f}}}~~.
\end{array}
\end{eqnarray}
Equation (\ref{ap:rho_dif}) is now solved by introducing a polynomial solution. 
A fifth order polynomial in $t$ is sufficient to satisfy the conditions of Eq. (\ref{ap:BC}). We substitute the polynomial solution into Eq. (\ref{ap:rho_dif}) and solve for $\omega\b t$.

To obtain the expectation values of $\hat{H},\,\hat{L}$ and $\hat{C}$, we introduce the eigenstates of $\hat{I}\b t$. These obey the eigenvalue equation
\begin{equation}
\hat{I}\b t \ket{\lambda \b t} = \lambda \ket{\lambda\b t}~~,     
\end{equation}
with time-dependent eigenstates and time-independent eigenvalues $\lambda$. Next, we define the annihilation operator  $\hat{c}={2\hbar}^{-1}\b{\hat{Q}/\rho-i\b{m\dot{\rho}\hat{Q}-\rho \hat{P}}}$ and matching annihilation operator. These operators satisfy:
$\hat{c}\ket{\lambda}=\sqrt{\lambda}\ket{\lambda-1}$ and $\hat{c}^\dagger \ket{\lambda} = \sqrt{1+\lambda}\ket{\lambda+1}$. By expressing $\hat{H},\,\hat{L}$ and $\hat{C}$ in terms of $\hat{c}$ and $\hat{c}^\dagger$ one obtains
\begin{eqnarray}
\begin{array}{ll}
 \bra{\lambda}\hat{H}\ket{\lambda} = \f{\hbar}{2} \b{m\dot{\rho}^2+\f{1}{m\rho^2}+m\omega^2\b t \rho^2}\b{\lambda+\f{1}{2}}  \\
  \bra{\lambda}\hat{L}\ket{\lambda} = \f{\hbar}{2} \b{m\dot{\rho}^2+\f{1}{m\rho^2}-m\omega^2\b t \rho^2}\b{\lambda+\f{1}{2}}\\
    \bra{\lambda}\hat{C}\ket{\lambda} = \omega\b t\hbar m\dot{\rho}\rho\b{\lambda+\f{1}{2}}~~.
\end{array}
\end{eqnarray}
The final step is to sum over the contribution of each state $\ket{\lambda}$. For an initial Gibbs state the energy reads
\begin{eqnarray}
\begin{array}{ll}
    \mean{\hat{H}}=
    \rm{tr}\b{\hat{\rho}\hat{H}}&=\sum_{\lambda,\lambda'}\rm{\ensuremath{\bra{\lambda}}}\hat{\rho}\ket{\lambda'}\rm{\ensuremath{\bra{\lambda'}}}\hat{H}\ket{\lambda}\\&=\sum_{\lambda}\rm{\ensuremath{\bra{\lambda}}}\hat{\rho}\ket{\lambda}\rm{\ensuremath{\bra{\lambda}}}\hat{H}\ket{\lambda}\\&
    =\sum_{\lambda}\f{e^{-\hbar\omega\b 0\b{\lambda+\f 12}/k_B T}}Z\rm{\ensuremath{\bra{\lambda}}}\hat{H}\ket{\lambda}\\&
    =\f 12\b{m\dot{\rho}^{2}+\f 1{m\rho^{2}}+m\omega^{2}\b t\rho^{2}}\f 12\rm{coth}\b{\f{\hbar\omega\b 0}{2k_{B}T}}~.
    \end{array}
\end{eqnarray}
The derivation for $\mean{\hat{L}}$ and $\mean{\hat{C}}$ follows a similar procedure.

\section{Friction action}
\label{sec:friction action}
Shortcut to equilibrium processes rely on non-adiabatic driving of an open quantum system. The driving incorporates the dissipative and unitary dynamics to lead the system towards a target thermal state. These protocols accelerate the system thermalization rate by generating coherence at intermediate times (transforming energy to coherence), and terminating them at the end of the protocol. The protocol duration $\tau_{stroke}$ of the STE can be varied within the inertial approximation, with a negligible influence on the final fidelity. For increasing protocol duration,  the generation of coherence reduces, converging to the quantum-adiabatic result in the limit $\tau_{stroke}\ra \infty$. As a result, as the protocol duration increases less coherence dissipates to the bath and the work output improves (reduced in the convention of $W<0$ for work extraction). Asymptotically, the work output scales  as $\tau_{stroke}^{-1}$ \cite{dann2018shortcut}, therefore, one can  introduce the `friction action' $F$ and express the total cycle work output $W$ as a function of the ideal work $W_{C}$ and $F$
\begin{equation}
    W= W_{C}+\f{F}{\tau_{cycle}}~~.
    \label{eq:W}
\end{equation}
In the quantum adiabatic limit the process is optimal and the work output converges to $W_{C}$, Eq. (\ref{eq:Work_ideal}).
Substituting Eq. (\ref{eq:W}) into the expression for the power output ${\cal{P}}=-W/\tau_{cycle}$, leads to the asymptotic relation
\begin{equation}
    {\cal{P}}-\f{|W_{C}|}{\tau_{cycle}}\propto -\f{F}{\tau_{cycle}^2}~~.
    \label{eq:P eq prop2}
\end{equation}
Equation (\ref{eq:P eq prop2}) is consistent with the general argument that the frictional forces should be independent of the sign of $\dot{\omega}$. Hence, to lowest order the power against friction is proportional to $\mu^2$. Since $\mu\propto1/\tau_{cycle}$ we expect that the assymptotic power against friction scales as Eq. (\ref{eq:P eq prop2}) \cite{francica2019role}.

\section{Endo-global cycle construction}
\label{apsec:Endo-global cycle}
The Endo-global cycle is constructed by combining two open strokes and two adiabats, that operate at a constant adiabatic speed. During the open strokes, the working medium dynamics is generated by the non-adiabatic master equation Eq. (\ref{eq:NAME}). This equation incorporates both the dissipative and unitary effects. It is exact for Markovian dynamics in the weak coupling regime and when the external driving is slow relative to the bath dynamics (see Ref. \cite{dann2018time}).  The condition of constant adiabatic speed, $|\mu|=\rm{const}$, leads to protocols of the following form $\omega \b t=\omega_i/\b{1-\omega_i \mu t}$, with initial frequency $\omega_i$. 

The unitary dynamics are given in terms of an operator basis in Liouville space, $\v v \b t=\{\hat{H}\b t$, $\hat{L}\b t,\hat{C}\b t, \hat{I}\}^T$.  Here, $\hat{H}\b t$ is the Hamiltonian, Eq. (\ref{eq:Ham}),
 $$\hat{L}\b t=\f{\hat{P^2}}{2m}-\f{1}{2}m\omega\b t^2\hat{Q}^2$$
 is the Lagrangian, and
 $$ \hat{C}\b t=\f{\omega\b t}{2}\b{\hat{Q}\hat{P}+\hat{P}\hat{Q}}$$
is the position-momentum correlation operator and $\hat{I}$ is the identity operator.
These operators completely determine the dynamics of the harmonic oscillator Gaussian state (\ref{eq:Gibbs state}), and give a clear physical interpretation of energy and coherence, 
\begin{equation}
    {\it{Coh}}=\f{\sqrt{\mean{\hat{L}\b t}^2+\mean{\hat{C}\b t}^2}}{\hbar \omega \b t}~~.
    \label{eq:coherence}
\end{equation}
The solution for $\v v\b t$  is given by \cite{kosloff2017quantum}:
\begin{equation}
    \f{d\v v}{dt} = {\cal{U}_S}\b t\v{v}\b 0~~,
\end{equation}
with
\begin{equation}
\label{eq:Adiprop}
{{\cal U}_S}\b t=\f 1{\kappa^2}\f{\omega\b t}{\omega\b 0}\left[{\begin{array}{cccc}
4-\mu^{2}c & -\mu\kappa s & -2\mu\b{c-1} & 0\\
-\mu\kappa s & \kappa^{2}c & -2\kappa s & 0\\
2\mu\b{c-1} & 2\kappa s & 4c-\mu^{2} & 0\\
0 & 0 & 0 &  {\kappa^2}\f{\omega\b 0}{\omega\b t}
\end{array}}\right]
\end{equation}
where $\kappa = \sqrt{4-\mu^2}$ and $c=\cos\b{\kappa \theta \b t}$, $s=\sin\b{\kappa \theta \b t}$.
The free propagation, governed by Eq. (\ref{eq:Adiprop}), mixes coherence and populations due to non-adiabatic driving \cite{kosloff2017quantum}.

In order to judge the performance for varying cycle-times, we construct different cycles with the same cycle frequencies, and vary $|\mu|$. The adiabatic parameter determines the stroke duration according to $\tau_{stroke} = \mu^{-1}\b{\omega_f -\omega_i}/\b{ \omega_f \omega_i }$, where $\omega_f$ is the final frequency. The cycles are then propagated until convergence to the  limit-cycle \cite{feldmann2004characteristics,insinga2018quantum},
where the performance is evaluated.

\section{Coherence measures}
\label{apsec:coherence operations}
Coherence is associated with non-diagonal elements in the energy representation. This means that a state possessing coherence is non-stationary under the free Hamiltonian dynamics. There have been many proposals to quantify coherence \cite{baumgratz2014quantifying}. In this context, two measures have been employed to analyze quantum heat engines: (i)  Divergence, which becomes the difference between the energy and  the von Neumann entropies \cite{lindblad1974expectations,uzdin2018global,feldmann2016transitions,camati2019coherence}, and (ii) The algebraic definition of coherence, utilized here, Eq. (\ref{eq:coherence}).



\section{Addition of pure dephasing}
\label{sec:dephasing}

Pure dephasing is introduced during the adiabats of the Endo-global cycle to evaluate the importance of coherence in the engine operation. Such dephasing can be caused by noise in the driving field and weak measurement \cite{feldmann2006quantum,levy2017action,konrad2010monitoring,wiseman2009quantum}. A Lindbladian describing dephasing is added to the free dynamics. 
In the Heisenberg picture the master equation becomes
\begin{eqnarray}
\begin{array}{ll}
   \f{d\hat{X}\b t}{dt}=&\f{i}{\hbar}\sb{\hat{H}\b t,\hat{X}\b t}+\pd{\hat{X}\b t}{t}-\gamma_d \sb{\hat{H}\b t,\sb{\hat{H}\b t,\hat{X}\b t}}~~.
   \end{array}
   \label{eq: deph term}
\end{eqnarray}
Here, the last term induces pure dephasing in the instantaneous energy basis, with a dephasing strength $\gamma_d$. We solve the dynamics, by representing the system in terms of the operator basis
 $\v v\b t=\{\hat{H}\b t$, $\hat{L}\b t,\hat{C}\b t, \hat{I}\}^T$, Cf.  \ref{apsec:Endo-global cycle}. Substituting the basis operators into Eq. (\ref{eq: deph term}) we express the dynamics of $\v v \b t$ in a matrix-vector notation
 \begin{equation}
     \f d{dt}\v v\b t=\omega\b t\b{\mu{\cal I}+{\cal M}}\v v\b t~~,
     \label{eq:dephasing dynamics}
 \end{equation}
 where $\mu$ is the adiabatic parameter, $\cal{I}$ is the identity matrix and $\cal M$ is given by 
 \begin{equation}
    {\cal{M}} =\sb{\begin{array}{cccc}
0 & -\mu & 0 & 0\\
-\mu & -4k_{d}\omega\b t & -2 & 0\\
0 & 2 & -4k_{d}\omega\b t & 0\\
0 & 0 & 0 & 1
\end{array}}~~.
 \end{equation}
 Equation (\ref{eq:dephasing dynamics}) is solved with a standard Runge-Kutta-Dormand-Prince propagator, leading to the  system dynamics in the presence of pure-dephasing.
 
 The reconstruction of the density operator, Eq. (\ref{eq:Gibbs state}), from the $\{\hat{H}\b t,  \hat{L}\b t,\hat{C}\b t,\hat{I}\}$ operator basis assumes that the working medium is described by a generalized canonical state. Strictly, the dephasing dynamics, Eq. (\ref{eq: deph term}), does not conserve the Gaussian structure. Nevertheless, for the cases studied, where the coherence is relatively small, the Gaussian state is a valid representation.

\section{Model parameters}
\label{apsec:model parameters}
The cycle parameters are described in Tables \ref{table:cycle parameters} and \ref{table:cycle parameters2}.
\begin{table}
\caption{Cycle parameters}
\label{table:cycle parameters2}
\begin{center}
\begin{tabular}{ |m{5cm}||m{5cm}|  }
 \hline
 Cycle parameter & Value (atomic units)\\
 \hline
 \hline
 Coupling strength, $\f{|\v d|^2}{4\pi \epsilon_0 \hbar c}$   & 0.05\\
 \hline
Stroke duration of the adiabatic expansion of the shortcut cycles& 5\\
 \hline
 Stroke duration of the adiabatic compression of the shortcut cycles& 5\\
 \hline
 
\end{tabular}
\end{center}
\end{table}
\clearpage
\section*{References}

\end{document}